\documentclass[aps,pra,preprintnumbers,showpacs,twocolumn,bibnotes,amsmath,amssymb,dvips]{revtex4}

\usepackage{graphicx}
\usepackage{bm}

\newcommand{\unit}[2]{{#1}~\ensuremath{\mathrm{#2}}}
\newcommand{\Ket}[1]{\ensuremath{|{#1}\rangle}}
\newcommand{\Bra}[1]{\ensuremath{\langle{#1}|}}

\begin{document}

\title{Rabi Spectroscopy and Excitation Inhomogeneity in a 1D Optical
  Lattice Clock}

\author{S. Blatt}
\email{sebastian.blatt@colorado.edu}
\author{J. W. Thomsen}
\altaffiliation{Permanent address: The Niels Bohr Institute,
  Universitetsparken 5, 2100 Copenhagen, Denmark}
\author{G. K. Campbell}
\author{A. D. Ludlow}
\altaffiliation{Current address: Time and Frequency Division, National
  Institute of Standards and Technology, 325 Broadway, Boulder,
  Colorado 80305, USA}
\author{M. D. Swallows}
\author{M. J. Martin}
\author{M. M. Boyd}
\altaffiliation{Current address: AOSense Inc., 767 N. Mary Ave, Sunnyvale,
  CA 94085-2909, USA}
\author{J. Ye}
\affiliation{JILA, National Institute of Standards and Technology and
  University of Colorado, \\
  Department of Physics, University of Colorado, Boulder, CO,
  80309-0440, USA}

\begin{abstract}
  We investigate the influence of atomic motion on precision Rabi
  spectroscopy of ultracold fermionic atoms confined in a deep, one
  dimensional (1D) optical lattice. We analyze the spectral components
  of longitudinal sideband spectra and present a model to extract
  information about the transverse motion and sample temperature from
  their structure. Rabi spectroscopy of the clock transition itself is
  also influenced by atomic motion in the weakly confined transverse
  directions of the optical lattice. By deriving Rabi flopping and
  Rabi lineshapes of the carrier transition, we obtain a model to
  quantify trap state dependent excitation inhomogeneities. The
  inhomogeneously excited ultracold fermions become distinguishable,
  which allows $s$-wave collisions. We derive a detailed model of this
  process and explain observed density shift data in terms of a
  dynamic mean field shift of the clock transition.
\end{abstract}

\pacs{37.10.Jk 42.50.Md 42.62.Eh 34.50.Cx}
\maketitle

\section{Introduction}
\label{sec:intro}

Evaluations of optical frequency standards based on thousands of
ultracold neutral $^{87}$Sr atoms confined in one dimensional (1D)
optical lattices now surpass~\cite{Ludlow08,Campbell08} the best
$^{133}$Cs primary standards~\cite{Heavner05,Bize05,Weyers01}. The
high precision achieved with the Sr optical lattice clock relies on
high quality factor optical spectroscopy~\cite{Boyd06}, enabled by
confining the atoms in a magic wavelength optical lattice~\cite{Ye08}
and probing along the strong confinement axis in the Lamb-Dicke and
resolved sideband regimes~\cite{Wineland79,Leibfried03}. No studies
yet indicate a fundamental limit at the $10^{-17}$ level of
accuracy\cite{Ludlow08,Campbell09,Baillard08,Akatsuka09}. Reaching new levels of accuracy demands
that ever more subtle spectroscopic effects are taken into account.
Recently, a density related clock frequency shift was
measured~\cite{Campbell09}, even though the clock is based on
fermionic $^{87}$Sr where collisions at temperatures of a few $\mu$K
are suppressed by the Pauli exclusion principle.

In this paper, we present a systematic experimental and theoretical
investigation of the Rabi spectroscopy process used in the $^{87}$Sr
optical lattice clock. This investigation results in a more detailed
model of the density-dependent clock frequency shift. These results
are used to describe our recent experimental work~\cite{Campbell09},
which has led to several different theoretical models of fermion clock
frequency shifts~\cite{Rey09,Gibble09}.

The experimental setup is described in Sec.~\ref{sec:setup}. We
present a perturbative model of the 1D lattice in
Sec.~\ref{sec:potential}, derive sideband spectra in
Sec.~\ref{sec:sideband} and show how temperature can be obtained from
their shape. Furthermore, we discuss time-dependent Rabi flopping and
spectroscopic lineshapes as well as their dependence on sample
temperature and probe-induced inhomogeneities in
Sec.~\ref{sec:carrier}. In particular, we identify the effect of
transverse motion on the excitation process as the main source of
inhomogeneity in the system, causing atoms to lose their
indistinguishability. Accurate modeling of the process in
Sec.~\ref{sec:density} allows relating this loss of
indistinguishability to a dynamic clock frequency shift proportional
to the atomic density.

\section{Experimental Setup}
\label{sec:setup}

\begin{figure}[htbp]
\includegraphics*[width=0.9\columnwidth]{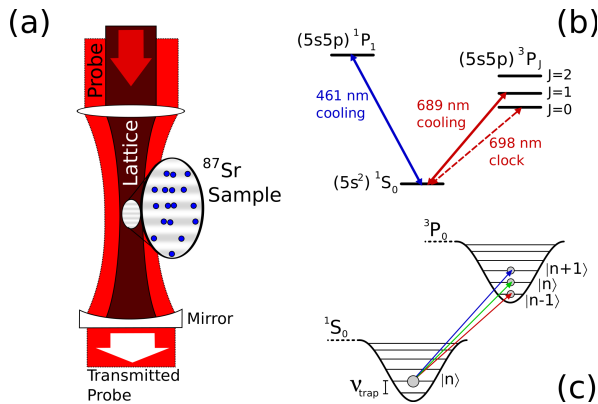}
\caption{(color online). (a) Experimental setup. A one dimensional
  optical lattice is oriented vertically along the $z$-axis to
  suppress tunneling. The lattice wavelength is \unit{813.43}{nm}. The
  clock probe beam is aligned collinear with the optical lattice, and
  the probe beam is transmitted by the mirror used to retro-reflect
  the lattice. (b) The optical transitions used to cool are
  ${^1S_0}-{^1P_1}$ at \unit{461}{nm}, and ${^1S_0}-{^3P_1}$ at
  \unit{689}{nm}. The spectroscopic transition between the ground
  $\Ket{g}$ and excited $\Ket{e}$ clock states is the doubly forbidden
  intercombination transition ${^1S_0}-{^3P_0}$ at \unit{698}{nm}. (c)
  High quality factor spectroscopy performed on lattice trapped atoms
  gives rise to three dominant spectral features; a central carrier
  (green arrow) where the motional state $\Ket{n}$ is conserved,
  accompanied by a red and blue sideband (red and blue arrows) where
  the motional state decreases and increases by one, respectively.}
\label{fig:setup}
\end{figure}

The ultracold atomic sample of $^{87}$Sr is produced by standard laser
cooling techniques and trapped at the antinodes of a vertically
oriented 1D optical lattice. The $^{87}$Sr ${^1S_0}-{^3P_0}$ clock
transition is subsequently interrogated with laser light propagating
collinear with the lattice axis. The geometry is shown schematically
in Fig.~\ref{fig:setup}(a). After the spectroscopic probe has
redistributed atomic population between the ground \Ket{g} ($^1S_0$)
and excited \Ket{e} ($^3P_0$) clock states, the populations are
measured via fluorescence detection on the ${^1S_0}-{^1P_1}$
transition, heating the sample out of the trap. By repeating many such
measurements, data is aggregated while the spectroscopy laser is
scanned across the clock transition to acquire a spectrum, the probe
pulse time is varied to obtain population dynamics, or the laser
frequency is stabilized to the clock transition for clock operation.

Details of the setup are described elsewhere~\cite{Boyd07b,Ludlow08b},
here we summarize the important experimental parameters for reference.
The atoms are cooled in a two stage magneto-optical trap (MOT) on the
transitions indicated by solid arrows in Fig.~\ref{fig:setup}(b). The
first stage uses the strong (\unit{30}{MHz}) ${^1S_0} - {^1P_1}$
transition at \unit{461}{nm}. The second stage MOT uses dual-frequency
narrow line cooling~\cite{Mukaiyama03,Loftus04} on the ${^1S_0}
(F=9/2)-{^3P_1} (F=9/2,11/2)$ intercombination lines (\unit{7}{kHz})
at \unit{689}{nm}. The optical lattice beam is overlapped with the
second stage MOT and atoms are directly cooled into the lattice. The
1D optical lattice is formed by two counterpropagating laser beams
near the Stark cancellation wavelength $\lambda=\unit{813.43}{nm}$,
where the differential polarizability of $^1S_0$ and $^3P_0$ clock
states is zero~\cite{Ye08}. In this experiment, a gaussian beam tilted
at a small angle with respect to gravity is focused to a waist
$w_0\simeq\unit{30}{\mu m}$ ($1/e^2$ radius of intensity) and
retroreflected by a spherical mirror with matching curvature. The
optical lattice forms in the overlap between the two beams. The
typical lattice depth is $U_0\simeq 130\times
h\nu_\text{rec}\simeq\unit{22}{\mu K}$, where $\nu_\text{rec} = h / (2
m \lambda^2)$ is the lattice recoil frequency.

The sample temperature is controlled by additional cooling (heating)
resulting in a sample of $\sim 4000$ atoms at temperatures variable
from 1 to \unit{5}{\mu K}. The atoms are Doppler cooled (heated) along
the transverse directions, and sideband cooling (heating) is performed
along the lattice axis with light detuned from the ${^1S_0} (F =
9/2)-{^3P_1} (F = 11/2)$ transition.

Using optical pumping, the atoms are then spin-polarized in a weak
magnetic field. By choosing the correct polarization for the optical
pumping beam, atoms are polarized into one of the two maximally
polarized states with magnetic quantum number $m_F = \pm 9/2$. During
the polarization step, the sideband cooling (heating) light is
simultaneously applied.

For carrier spectroscopy relevant to optical clock operation we probe
the ${^1S_0} - {^3P_0}$ clock transition along the strong confinement
axis of the lattice in the Lamb-Dicke regime and the resolved sideband
limit~\cite{Wineland79,Leibfried03} with a sub-Hertz cavity-stabilized
diode laser~\cite{Ludlow07}. Here, the trap frequency of
\unit{70-80}{kHz} is much larger than the probe recoil frequency of
\unit{4.7}{kHz} and the natural transition linewidth of
\unit{1}{mHz}~\cite{Boyd07,Porsev04,Santra04}. This ensures a clock
interrogation which is highly insensitive to Doppler and recoil
effects. The duration of a clock spectroscopy pulse is \unit{80}{ms}
and the probe beam has a $\sim$5 times larger waist than the
lattice beam to minimize inhomogeneities introduced by the gaussian
probe beam profile. To probe time-domain Rabi flopping, the pulse time
is varied and the motional sidebands are probed by detuning the clock
laser by the motional trap frequency.

After applying the spectroscopy pulse, atoms remaining in the ground
state are detected by measuring fluorescence on the ${^1S_0} -
{^1P_1}$ transition. This detection pulse heats all $^1S_0$ atoms out
of the trap. The population in the $^3P_0$ state is then measured by
first pumping the atoms back to the $^1S_0$ state via $^3P_1$ and by
again measuring fluorescence on the \unit{461}{nm} transition.
Combining data from these two measurements results in a normalized
excitation fraction. The optical lattice is reloaded with a new sample
for each new measurement with a $\sim$\unit{1}{s} cycle time.

\section{Spectroscopy in a one-dimensional optical lattice}
\label{sec:potential}

Neglecting optical aberrations in the transverse beam profile, the
resulting trapping potential near the waist can be written as
\begin{equation}
  \label{eq:1}
  U(z,r) = - U_0 \cos^2(kz)~e^{-2r^2/w_0^2} + m g z,
\end{equation}
where $r=\sqrt{x^2+y^2}$ designates the transverse distance from the
lattice axis, $z$ is the longitudinal coordinate, $k=2\pi/\lambda$ is
the lattice wavenumber, $g\simeq\unit{9.81}{m/s^2}$ is the gravitational
acceleration, and $m$ is the mass of $^{87}$Sr. The resulting trap
is a nearly vertical stack of flat ellipsoids (``pancakes'') with an
aspect ratio given by lattice wavelength and beam waist.

The gravitational energy shift between neighboring pancakes
($\sim\unit{1}{kHz}$) breaks the translational symmetry of the
potential and for our trap depths intersite tunneling becomes strongly
suppressed~\cite{Lemonde03}. The resulting separation of a lattice
band into isolated sites is called a Wannier-Stark
ladder~\cite{Glueck02}. Each site has several states whose energies
are determined by the gravity-free lattice structure, but their energy
widths decrease according to the residual Landau-Zener tunneling
rates to neighboring sites. For our trap depths, temperatures of a few
$\mu$K, and low site populations, this complex problem simplifies
drastically: we can consider single particles in isolated sites,
neglect intersite tunneling, and use thermal averaging to evaluate
the relevant spectroscopic parameters.

However, even the single site potential is not separable into
independent coordinates and it has no analytical solutions. We
approximate the longitudinal potential in a site as a one-dimensional
harmonic oscillator in $z$ with a quartic distortion. The shallow
transverse potential given by the gaussian lattice beam profile is
approximated as a two-dimensional harmonic oscillator in $r$. We also
include the first order coupling term between the longitudinal and
transverse degrees of freedom and find
\begin{equation}
  \label{eq:2}
  U(z,r) \simeq U_0(-1 + k^2 z^2 + \frac{2}{w_0^2} r^2
  - \frac{k^4}{3}z^4 - \frac{2k^2}{w_0^2} z^2 r^2).
\end{equation}
Treating the quartic distortion and the coupling term in first order
perturbation theory for harmonic oscillator states
$\Ket{\bm{n}}=\Ket{n_x,n_y,n_z}$ gives an energy spectrum
\begin{equation}
  \label{eq:13}
  \begin{split}
  E_{\bm{n}}/h \simeq& \nu_z (n_z+\frac{1}{2}) + \nu_r (n_x+n_y+1)\\
  &-\frac{\nu_\text{rec}}{2}(n_z^2+n_z+\frac{1}{2})\\
  &-\nu_\text{rec}\frac{\nu_r}{\nu_z}(n_x+n_y+1)(n_z+\frac{1}{2}),
  \end{split}
\end{equation}
where we identify the longitudinal and transverse trap
frequencies from the harmonic approximation to the potential as
\begin{align}
  \label{eq:3}
  \nu_z &= 2\nu_\text{rec}\sqrt{\frac{U_0}{h\nu_\text{rec}}}\\
  \nu_r &= \sqrt{\frac{U_0}{m \pi^2 w_0^2}}.
\end{align}
These relations allow determination of the trap depth and beam waist
by measuring the trap frequencies. The number of states in the trap is
approximately given by $N_z N_r^2$, with
\begin{align}
  \label{eq:4}
  N_z &\simeq \frac{U_0}{h\nu_z} = \sqrt{\frac{U_0}{4h\nu_\text{rec}}} \\
  N_r &\simeq N_z \frac{\nu_z}{\nu_r}.
\end{align}
Typical longitudinal and transverse trap frequencies in our
experiment are \unit{80}{kHz} and \unit{450}{Hz}, respectively, so
that $N_z~(N_r) \simeq 6~(1000)$. Due to the quartic distortion by the
sinusoidal lattice potential, the longitudinal energy gap
\begin{equation}
  \label{eq:15}
  \begin{split}
  \gamma(n_z) &\equiv (E_{n_x,n_y,n_z+1}-E_{n_x,n_y,n_z})/h \\
  & = \nu_z - \nu_\text{rec}(n_z+1)- \nu_\text{rec} \frac{\nu_r}{\nu_z}(n_x+n_y+1),
  \end{split}
\end{equation}
determines the measured longitudinal trap frequency as approximately
$\nu_z-\nu_\text{rec}$ instead of $\nu_z$. In contrast, the $r^2z^2$
coupling term in Eqn.~\ref{eq:2} has a more subtle effect. Radially
oscillating atoms explore trap regions with different longitudinal
trap frequencies and their response to the spectroscopy laser changes.

Vibrational laser spectroscopy of trapped atoms is a well described
topic~\cite{Wineland79,Leibfried03}. The Rabi frequency associated
with a traveling wave probe of wave vector $\bm{k}_p$ between initial
and final trap states $\Ket{\bm{n}_i}$ and $\Ket{\bm{n}_f}$ is
\begin{equation}
  \label{eq:6}
  \Omega_{\bm{n}_f\leftarrow \bm{n}_i} =
  \Omega_0 \Bra{\bm{n}_f} e^{i \bm{k}_p\cdot\bm{x}} \Ket{\bm{n}_i},
\end{equation}
where the free-space Rabi frequency $\Omega_0$ is given by the dipole
matrix element between the electronic states. Thus atoms in different
motional states will respond differently to the same probe field. The
resulting spread in Rabi frequencies leads to inhomogeneities in the
system for any non-zero temperature.

In the following sections, we will model the most prominent
spectroscopic features in a scan across the carrier and motional
sideband resonances and compare these models to experimental data to
extract information about sample temperatures. The temperature as well
as an effective probe laser misalignment will then be used to quantify
the degree of inhomogeneity in the system, leading to a model of
collisional frequency shifts for ultracold fermions in a 1D optical
lattice.

\section{Sideband Spectroscopy}
\label{sec:sideband}

\begin{figure}[htbp]
\includegraphics*[width=0.9\columnwidth]{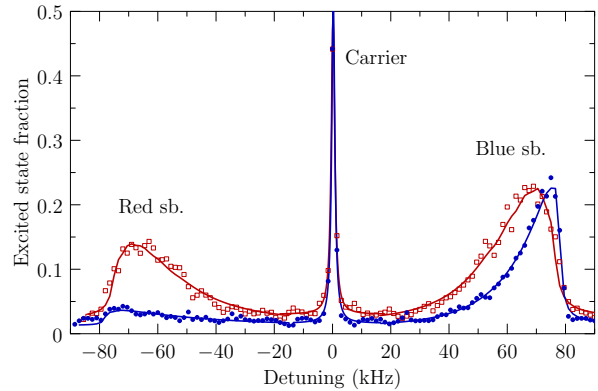}
\caption{(color online). Sideband spectra with longitudinal and
  transverse cooling (circles) and without in-trap cooling (open
  squares) by scanning the spectroscopy laser across the electronic
  transition with a Rabi pulse time of \unit{80}{ms} and Rabi
  frequencies on the order of \unit{1}{kHz}. The solid lines are
  combinations of fits to the carrier and sidebands.}
\label{sidebandscan}
\end{figure}

The motional sideband spectrum of lattice-trapped atoms exhibits
features similar to the well known spectra of harmonically trapped
ions~\cite{Leibfried03}. Each data point in Fig.~\ref{sidebandscan}
was obtained by measuring the excited state fraction after choosing a
probe laser detuning in the vicinity of the clock transition and
applying an \unit{80}{ms} Rabi pulse. The traces show a narrow carrier
transition that is (mostly) free of motional effects, accompanied by
motional sidebands at the longitudinal trap frequency
($\nu_z\simeq\unit{80}{kHz}$). The red-detuned sideband
($n_z\rightarrow n_z-1$) is suppressed with respect to the
blue-detuned sideband ($n_z\rightarrow n_z+1$), indicative of the
temperature along the strong confinement axis. In contrast to ion trap
experiments, the sidebands are smeared out towards the
carrier~\cite{Ludlow06}. This skewing is due to the coupling between
the longitudinal and transverse degrees of freedom which makes the
longitudinal transition frequency depend on the transverse motional
state. The resulting line shape carries information about the
potential and the longitudinal and transverse temperatures of the
trapped atoms.

The clock transition's natural linewidth is on the order of mHz
corresponding to a metastable lifetime on the order of
\unit{150}{s}~\cite{Boyd07}. At large probe powers, saturation
broadening can increase the carrier linewidth to $\sim$\unit{1}{kHz}.
At typical Rabi probe times of \unit{80}{ms}, the population dynamics
have dephased to an equilibrium (see Sec.~\ref{sec:carrier}), allowing
a time-independent model of the sideband shape. In the following, we
will focus on the first blue longitudinal sideband.

The blue longitudinal sideband is produced by atoms undergoing the
clock transition along with a motional transition $\Ket{\bm{n}_i}
=\Ket{n_x,n_y,n_z} \rightarrow \Ket{\bm{n}_f}=\Ket{n_x,n_y,n_z+1}$.
The probe laser detuning $\delta$ exciting this transition is given by
the energy difference in Eqn.~\ref{eq:15} and depends on the
transverse motional state. The relative blue sideband amplitude for a
given probe detuning is thus given by the transverse motional
distribution. By assuming a thermal distribution among the transverse
trap states, we can relate the shape of the shallow sideband edge
(facing the carrier) to the transverse temperature, as shown in
Appendix~\ref{sec:lineshape-deriv}. We find an approximate line shape
as a function of detuning $\delta$ from the carrier (valid for the
shallow sideband edge):
\begin{equation}
  \label{eq:11}
  \begin{split}
  \sigma^{n_z}_\text{blue}(\delta) &=
  \frac{\alpha^2}{\tilde\gamma(n_z)}\left[1-\frac{\delta}{\tilde\gamma(n_z)}\right]
  e^{-\alpha\left[1-\delta/\tilde\gamma(n_z)\right]}\\
  &\times\Theta[\tilde\gamma(n_z)-\delta],
  \end{split}
\end{equation}
with $\alpha =
\frac{\tilde\gamma(n_z)}{\nu_\text{rec}}\frac{h\nu_z}{k_B T_r}$,
uncoupled longitudinal energy gap $\tilde\gamma(n_z) =
\nu_z-\nu_\text{rec}(n_z+1)$, and the Heaviside function $\Theta$. The
sideband shape in Eq.~\ref{eq:11} is approximately area-normalized and its
main feature is an exponential slope falling towards the
carrier, produced by the changing longitudinal energy gap with
transverse excursion (the third term in Eqn.~\ref{eq:15}). This
exponential is multiplied by a linear term rising towards the
carrier and vanishing at $\delta=\tilde\gamma(n_z)$, which arises from
the degenerate two-dimensional transverse confinement. These two
features dominate the shape of the sideband and capture the shallow
slope towards the carrier. Equation~\ref{eq:11} neglects
the transition between the shallow edge and the underlying
power-broadened Lorentzian that determines the sharp edge.

The final, thermally averaged, blue sideband absorption cross section
is then given by (see Appendix~\ref{sec:lineshape-deriv})
\begin{equation}
  \label{eq:12}
  \sigma_\text{blue}(\delta) \propto \sum_{n_z=0}^{N_z}
  e^{-\frac{E_{n_z}}{k_B T_z}}
  \sigma^{n_z}_\text{blue}(\delta),
\end{equation}
as a Boltzmann-weighted superposition of single (longitudinal) state
sidebands shifted by the anharmonicity of the longitudinal trap. Here
$E_{n_z}$ is the energy of longitudinal state $n_z$, neglecting the
$r^2z^2$ coupling term. In addition, each $n_z\rightarrow n_z+1$
sideband is smeared out towards the carrier by the coupling term
between the longitudinal and transverse traps. Each component's base
Lorentzian would additionally be broadened by the lifetime of the
corresponding Wannier-Stark state. However, for the relevant states
and trap depths, this broadening is much smaller than the
power-broadened width and can be ignored.

Regardless of the details of the component lineshapes in
Eqn.~\ref{eq:12}, the only difference between the red and blue
sidebands should be that the Boltzmann weights are shifted according
to $n_z\mapsto n_z+1$ since the particle starts in the higher-lying
motional state. There is no contribution from the longitudinal ground
state to the red sideband cross section and the ratio of integrated
sideband absorption cross sections obeys
\begin{equation}
  \label{eq:17}
  \frac{\sigma^\text{total}_\text{red}}{\sigma^\text{total}_\text{blue}}
  = \frac{\sum_{n_z=1}^{N_z}e^{-\frac{E_{n_z}}{k_B T_z}}}
  {\sum_{n_z=0}^{N_z}{e^{-\frac{E_{n_z}}{k_B T_z}}}}
  =  1 - \frac{e^{-\frac{E_{0}}{k_B T_z}}}
  {\sum_{n_z=0}^{N_z}{e^{-\frac{E_{n_z}}{k_B T_z}}}},
\end{equation}
which can be solved numerically for $T_z$ after determining $\nu_z$
from the sharp sideband edge on the far side of the carrier, resulting
in a lineshape-independent measure of the longitudinal temperature.

Ideally, the transverse temperature $T_r$ would be determined with the
same method by measuring the transverse sidebands at $\delta=\nu_r$,
but their amplitude is suppressed since the probe is carefully aligned
with the lattice axis. The transverse sidebands can be measured by
misaligning the probe beam. An independent measurement of the trap
frequency can be obtained by parametric heating. Both methods confirm
a typical transverse trap frequency of $\nu_r\simeq\unit{450}{Hz}$. A
combination of a longitudinal sideband spectrum with a transverse trap
frequency measurement allows a complete characterization of the trap
parameters including depth $U_0$ and waist $w_0$ via Eqn.~\ref{eq:3}.
This calibration, together with a fit of the longitudinal sideband
shape with Eqn.~\ref{eq:12}, then lets us extract information about
$T_r$ from the same data that gives $\nu_z$ and $T_z$.

\begin{figure}[htbp]
  \centering
  \includegraphics[width=\columnwidth]{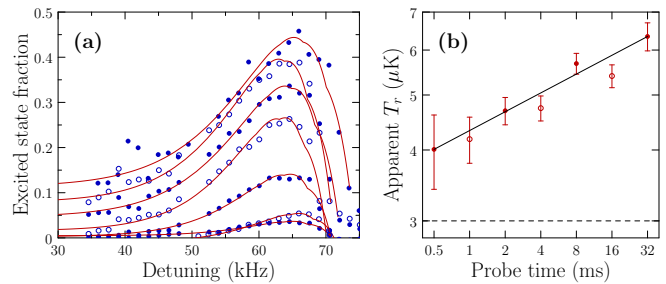}
  \caption{(color online). (a) Blue sideband as a function of probe
    time. The solid curves are fits of Eqn.~\ref{eq:12} to the data
    for fixed $T_z=\unit{3}{\mu K}$ determined from the cross section
    ratio in Eqn.~\ref{eq:17}. (b) Apparent transverse temperature
    as determined from the fits in (a) as a function of probe time on
    a double-logarithmic plot. The apparent transverse temperature
    increases with probe time, although unity-aspect TOF data
    indicates $T_r=T_z$ (dashed line). The solid line connects the
    first and last point, giving an increase in apparent transverse
    temperature as $\sim$\unit{1.5}{\mu K} per decade of probe time.}
  \label{fig:temp-vs-probetime}
\end{figure}

Spectroscopic lineshapes of the blue sideband are displayed in
Fig.~\ref{fig:temp-vs-probetime}(a), along with fits of
Eqn.~\ref{eq:12}, for varying Rabi pulse time in the range of
\unit{0.5-32}{ms}. The longitudinal temperature was determined by the
ratio of red and blue sideband areas (Eqn.~\ref{eq:17}) as
$T_z=\unit{3}{\mu K}$ and time-of-flight (TOF) pictures show a unity
aspect ratio, indicating a well thermalized sample with $T_z = T_r$.
Figure~\ref{fig:temp-vs-probetime}(b) shows the transverse
temperature resulting from the fits in (a) and the dashed line
indicates $T_z$; the solid line connects the first and last points.

The data show an apparent increase in transverse temperature with
increasing probe time. A similar measurement with constant pulse time
and varying probe power shows no such dependence. In combination with
the TOF measurements, we conclude that the transverse temperature is
not influenced by the probe laser, and that the magnitude of the Rabi
frequency does not influence the sideband shape except by broadening
the underlying Lorentzian. The time-dependent data suggest that the
transverse atomic motion lets more atoms interact with the probe laser
than suggested by the steady-state eigenenergy picture that gives rise
to Eqn.~\ref{eq:11}. With probe pulse times on the order of the
inverse trap frequency, a snapshot of the atomic distribution can be
obtained and the sideband shapes derived above become more applicable.

\section{Effect of radial motion on the carrier transition}
\label{sec:carrier}

The longitudinal sideband shape is strongly modified by the anharmonic
corrections to the potential. In contrast we expect the carrier to be
insensitive to motional effects since carrier spectroscopy can be done
at line widths of $\sim$\unit{2}{Hz}, corresponding to spectroscopic
quality factors of $2\times 10^{14}$~\cite{Boyd06}, which is well in
the resolved sideband regime.

The carrier Rabi frequency for a harmonically trapped particle in
state $\Ket{\bm{n}}$, probed with a spectroscopy beam of wave vector
$\bm{k}_p = 2\pi/\lambda_p \hat{\bm{k}}_p$ for probe wavelength
$\lambda_p$ is given by
\begin{equation}
  \label{eq:18}
  \Omega_{\bm{n}} = \Omega_0
  \Bra{\bm{n}}e^{i\bm{k}_p\cdot\bm{x}}\Ket{\bm{n}}
= \Omega_0 \prod_{i=x,y,z} e^{-\eta_i^2/2} L_{n_i}(\eta_i^2),
\end{equation}
determined by the Lamb-Dicke parameters $\eta_i \equiv k^i_p a_i /
\sqrt{2}$, with oscillator length $2\pi a_i = \sqrt{h/(m\nu_i)}$ for
trap axis $i$ and Laguerre polynomial $L_{n_i}$. In this way, the Rabi
frequency is determined by the spatial extent of the motional wave
function with respect to the probe wavelength. For a probe that is
perfectly aligned with the lattice axis, the only non-zero Lamb-Dicke
parameter will be $\eta_z$. In this case, $\eta_z^2 =
\nu^p_\text{rec}/\nu_z$, where $\nu^p_\text{rec}\simeq\unit{4.7}{kHz}$
is the probe recoil frequency.

We ignore corrections to Eqn.~\ref{eq:18} from perturbed harmonic
oscillator states, since both the $z^4$ and $r^2z^2$ terms in
Eqn.~\ref{eq:2} produce wave function coefficients suppressed by
$\nu_\text{rec}/(2\nu_z)$. Although the $r^2z^2$ coupling modifies
$\eta_z$, the correction to its value scales as
$n_r\times\sqrt{\nu_\text{rec}\nu_r}/\nu_z$ and is negligible for the relevant
temperatures. In the following, we will assume harmonic oscillator
states and energies and derive carrier lineshapes and Rabi flopping
accordingly.

Although we can ignore the trap shape except for the harmonic
confinement, the Rabi frequency expression still assumes that the
particle is probed by a plane wave without transverse profile. Each
wave vector component of the probe beam contributes to the Rabi
frequency according to its Fourier coefficient, allowing us to
estimate the effect of a shaped probe beam. Over the extent of a trap
site the probe beam shape can be approximated as a plane wave along
the mean probe direction with a transverse shape function describing
its intensity profile. Although the probe beam is carefully aligned
with the lattice axis, residual misalignment or aberrations might
produce a net mismatch between lattice axis $\bm{k}=k\hat{\bm{z}}$ and
probe axis. Since the transverse trap is isotropic, we choose a small
net misalignment angle $\Delta\theta$ along $\hat{\bm{x}}$, such that
$\bm{k}_p \simeq k_p (\hat{\bm{z}} + \Delta\theta\hat{\bm{x}})$. The
transverse extent of the probe beam is large and can be approximated
by a cylindrically symmetric function (with respect to the net probe
direction) of waist $w_p \gg w_0$. The corresponding Rabi frequency is
\begin{equation}
  \label{eq:7}
  \begin{split}
  \Omega_{n_x,n_z} &= \Omega_0 \Bra{\bm{n}} e^{i\bm{k}_p\cdot\bm{x}}
  \left[1 +
    \mathcal{O}(a^2_x/w^2_p)\right]\Ket{\bm{n}}\\
  &\simeq \Omega_0
  \Bra{\bm{n}}e^{i\bm{k}_p\cdot\bm{x}}\Ket{\bm{n}}\\
  &= \Omega_0 e^{-\eta_x^2/2} e^{-\eta_z^2/2} L_{n_x}(\eta_x^2)
  L_{n_z}(\eta_z^2).
  \end{split}
\end{equation}
Thus the transverse shape of a cylindrically symmetric probe beam with
large cross section cannot influence the spectroscopy ($a^2_x/w^2_p
\simeq 10^{-4}$ for our experiment). The Lamb-Dicke parameters are
\begin{align}
  \label{eq:8}
  \eta_z &= 1/\lambda_p \sqrt{h/(2m\nu_z)}\\
  \eta_x &= \Delta\theta/\lambda_p \sqrt{h/(2m\nu_r)}
\end{align}
Since $\nu_r\ll\nu_z$, even a small amount of effective misalignment
will cause significant transverse contributions to the carrier
lineshape. For our trap frequencies, $\eta_z \simeq 0.24$, and $\eta_x
\simeq \Delta\theta\times 3.2$. In conclusion, the main correction to
the carrier Rabi frequency comes from an effective misalignment angle
in the transverse direction, while the broad transverse profile can be
ignored. This misalignment introduces information about the
transverse motional state distribution and thus the transverse
temperature into the carrier line shape. We will present experimental
evidence for a typical effective misalignment angle $\Delta\theta \simeq
\unit{10}{mrad}$.

The carrier line shape for Rabi spectroscopy can be understood as
follows. We can neglect spontaneous emission and probe laser
decoherence even at Rabi spectroscopy times approaching
\unit{1}{s}~\cite{Boyd06}. The lattice lifetime has also been measured
as $\simeq\unit{1}{s}$ and does not introduce decoherence on the
typical Rabi pulse times of \unit{80}{ms}. In this regime, we can
neglect any decoherence rates in the system and describe the
population dynamics in a fully coherent way. The excited state
probability for motional state $\Ket{\bm{n}}$, detuning $\delta$ and
pulse time $t$ is
\begin{equation}
  \label{eq:14}
  p_e(\bm{n},\delta,t) = \frac{\Omega^2_{n_x,n_z}}{\Omega^2_{n_x,n_z} + \delta^2}
  \sin^2\left[\pi t \sqrt{\Omega^2_{n_x,n_z} + \delta^2}\right].
\end{equation}
The ensemble-averaged excited state population is then given by
\begin{equation}
  \label{eq:16}
  P_e(\delta,t) = \sum_{n_x,n_z} q_{n_x}(T_r)
  q_{n_z}(T_z) p_e(\bm{n},\delta,t),
\end{equation}
for normalized Boltzmann-weights $q_{n_x}$ ($q_{n_z}$) corresponding
to transverse (longitudinal) temperature $T_r$ ($T_z$):
\begin{align}
  \label{eq:21}
  q_{n_x} &= (1-z_x) z_x^{n_x} & z_x\equiv\exp[-h\nu_r/(k_B T_r)]\\
  q_{n_z} &= (1-z_z) z_z^{n_z} & z_z\equiv\exp[-h\nu_z/(k_B T_z)]
\end{align}
We are interested in two scenarios: (a) Change the pulse time $t$ and
measure Rabi flopping at zero detuning. (b) Change detuning $\delta$
by scanning the spectroscopy laser across the carrier resonance and
measure the Rabi line shape. Both cases are covered by
Eqn.~\ref{eq:16} and we expect a response as a coherent superposition
of slightly different Rabi frequencies.

For low temperatures, we only have to consider atoms in the lowest
longitudinal state $n_z=0$ and for small misalignments, the Rabi
frequency can be expanded in $\eta_x^2$. These two
simplifications allow finding an analytical expression for case (a).
Here, the excited state population can be approximated as
\begin{equation}
  \label{eq:19}
  \begin{split}
  P_e(t) &\simeq \sum_{n_x=0}^{\infty} (1-z_x)z_x^{n_x}
  \sin^2[\phi(1-\eta_x^2 n_x)/2]\\
  & = \frac{1}{2} +
  \frac{1-z_x}{2}\frac{z_x\cos[\phi(1-\eta_x^2)]-\cos\phi}
  {1+z_x^2-2z_x\cos(\phi\eta_x^2)},
  \end{split}
\end{equation}
with $\phi = 2\pi t \Omega_0 e^{-\eta_x^2/2}e^{-\eta_z^2/2}$.
Equation~\ref{eq:19} reduces to $P_e(t) = \sin^2{\phi/2}$ for zero
misalignment and exhibits dephasing to $1/2$ by introducing different
frequency components via $\eta_x^2(\Delta\theta)$ and amplifying their
contribution by increasing $z_x(T_r)$. For case (b), the sum can be
evaluated in a similar manner. The resulting expression
does not provide much further insight and we omit it here.

Experimentally, the excited state fraction is obtained by measuring
the number of atoms left in the ground state after the spectroscopy
pulse, then repumping the excited atoms back to the ground state and
measuring their number. The repumping back to the ground state has
efficiency $\beta \le 1$ and can vary from day to day. To obtain an
excited state probability independent of a fluctuating overall number
of atoms, the fraction of excited state over excited plus ground state
counts is used. This normalization results in a Rabi flopping trace
that dephases to $\beta /( \beta + 1)$ instead of $1/2$. Note that on
the experimental timescale, the dephasing is a coherent process
dominated by an inhomogeneous distribution of Rabi frequencies via
effective probe misalignment and temperature.

\begin{figure}[htbp]
\includegraphics*[width=\columnwidth]{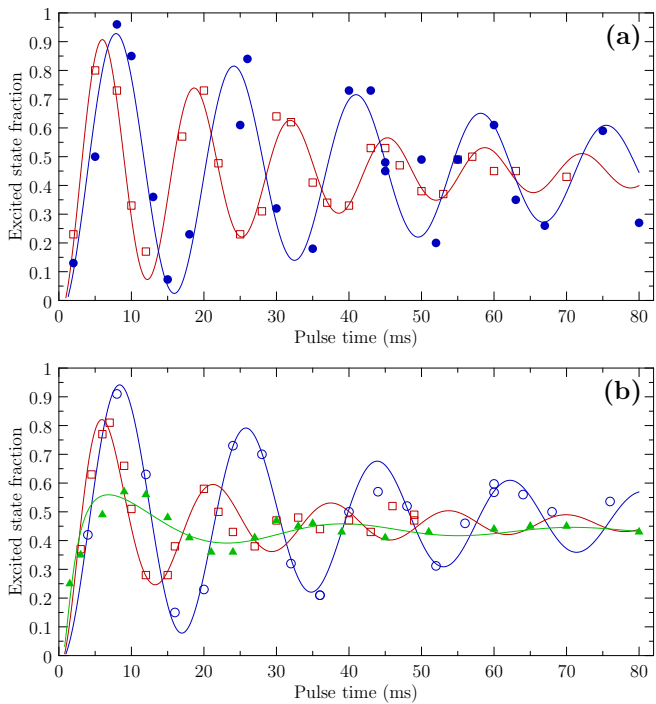}
\caption{(color online). Effect of inhomogeneity on Rabi oscillations.
  (a) The excited state fraction as a function of pulse time is shown
  for sample temperatures of \unit{1}{\mu K} (circles) and
  \unit{3}{\mu K} (open squares) with fits of Eqn.~\ref{eq:19}, giving
  bare Rabi frequencies $\Omega_0=\unit{59}{Hz}$ and \unit{76}{Hz},
  respectively. Both fits result in a misalignment angle
  $\Delta\theta\simeq\unit{10}{mrad}$. (b) The excited state fraction
  as a function of pulse time at \unit{3}{\mu K} is shown for
  increased misalignment angles and probe powers. The open circles
  show oscillations when the probe beam is slightly misaligned. The
  fit gives $\Omega_0=\unit{55}{Hz}$ and
  $\Delta\theta\simeq\unit{10}{mrad}$. The open squares show data for
  increased misalignment and the same probe power, giving $\Omega_0
  =\unit{54}{Hz}$ and $\Delta\theta\simeq\unit{17}{mrad}$. Note that
  the small misalignment approximation used to derive Eqn.~\ref{eq:19}
  starts to break down, resulting in a worse fit. Finally, the
  triangles show data for a large misalignment angle, taken with
  increased power. Although the fit looks much worse, the resulting
  $\Delta\theta\simeq\unit{40}{mrad}$ agrees well with a geometrical
  estimate based on the experimental procedure for misaligning the
  beam.}
\label{fig:rabiflopping}
\end{figure}

We investigated the effect of inhomogeneous excitation by mapping out
the Rabi flopping and lineshapes under different experimental
conditions such as sample temperature, misalignment angle and probe
laser intensity. In Fig.~\ref{fig:rabiflopping}(a) we show data for
the excited state fraction as a function of probe pulse time for two
different temperatures. For the hot sample, Rabi oscillations quickly
decay after only 2-3 cycles. For the cold sample, the Rabi
oscillations have a significantly higher visibility and are
observable for about 10 cycles. This behavior was reproduced for a wide
range of probe laser intensities. In Fig.~\ref{fig:rabiflopping}(b),
the Rabi oscillations dephase faster when the effective misalignment
angle $\Delta\theta$ is increased by misaligning the probe with
respect to the lattice axis. From fits of Eqn.~\ref{eq:19} we determine
a misalignment angle of $\Delta\theta = \unit{10}{mrad}$ for
the well-aligned case relevant to clock operation.

\begin{figure}[htbp]
\includegraphics*[width=\columnwidth]{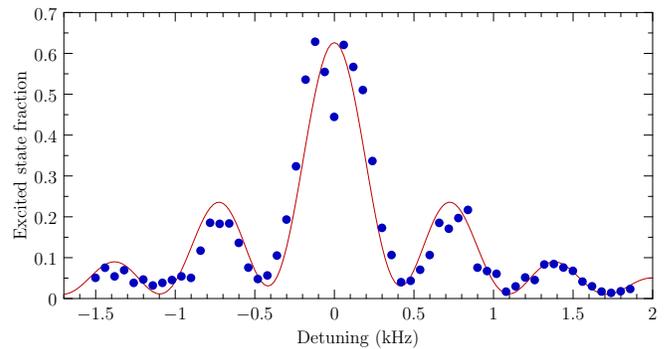}
\caption{(color online). Excitation fraction as a function of probe
  laser detuning for a probe time of \unit{1.7}{ms}. The solid curve
  is a fit of Eqn.~\ref{eq:16} with the temperature as the only free
  parameter, fixing $\Delta\theta$ at \unit{10}{mrad}. The fit gives
  $T_z = T_r = \unit{(2.1\pm 0.2)}{\mu K}$ consistent with the
  sideband method and time-of-flight expansion.}
\label{fig:rabitrace}
\end{figure}

Each data point in Fig.~\ref{fig:rabiflopping} is determined by
setting a specific probe pulse time and then scanning the probe laser
detuning across the carrier transition to obtain a spectrum such as
the one shown in Fig.~\ref{fig:rabitrace} for a probe time of
\unit{1.7}{ms}. The maximum excited state fraction at zero laser
detuning ($\sim 0.65$ here) would then give the corresponding data point in
Fig.~\ref{fig:rabiflopping} for the respective probe time.

For small $\Delta\theta$, the data shown in
Figs.~\ref{fig:rabiflopping} and~\ref{fig:rabitrace} are well
reproduced by the simplified model in Eqn.~\ref{eq:19}. This model
allows us to determine the transverse temperature as \unit{1}{\mu K}
for the cooled case, and \unit{3}{\mu K} for the uncooled case, with a
precision of about \unit{0.2}{\mu K}, agreeing well with the
corresponding longitudinal temperatures. In the misalignment model,
both angle and transverse temperature produce a similar effect so that
their covariance is significant. The transverse temperature data has
been confirmed by the unity aspect ratio of the atomic cloud in
time-of-flight expansion. Accurate determination of the effective
misalignment is difficult, introducing larger uncertainty in the
radial temperature measurement. However, fitting both Rabi flopping at
zero detuning and Rabi lineshapes across a wide range of parameters
gives consistent results.

\begin{figure}[htbp]
\includegraphics*[width=\columnwidth]{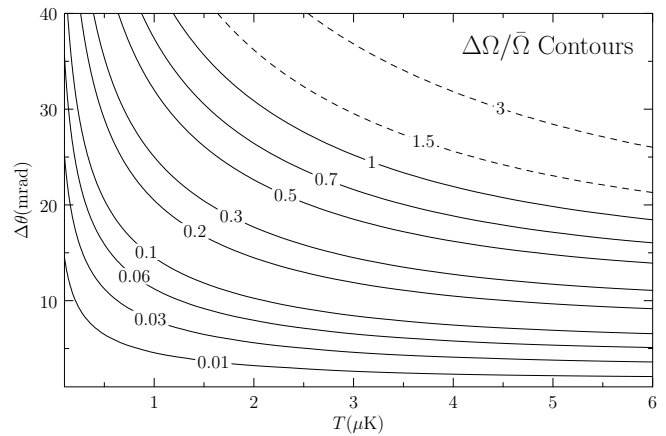}
\caption{(color online). Contour plot of the relative spread in Rabi
  frequency $\Delta\Omega/\bar\Omega$ as a function of misalignment
  angle $\Delta\theta$ and temperature $T_z=T_r=T$. The dashed contour
  lines indicate points where the mean Rabi frequency becomes smaller
  than the RMS spread, made possible by the negative values of the
  Laguerre polynomials in Eqn.~\ref{eq:7} for large radial quantum
  numbers. The experimentally relevant range is
  $\Delta\Omega/\bar\Omega = 0.05-0.4$.}
\label{fig:rabispread}
\end{figure}

We characterize the amount of inhomogeneity in the system by the ratio
of RMS spread in Rabi frequency $\Delta\Omega$ over the site's mean
Rabi frequency $\bar{\Omega}$, given by
\begin{align}
  \label{eq:22}
  \bar\Omega &= \sum_{n_x,n_z} q(n_x)q(n_z) \Omega_{n_x,n_z}\\
  \Delta\Omega^2 &= \sum_{n_x,n_z} q(n_x)q(n_z) \Omega^2_{n_x,n_z} - {\bar\Omega}^2.
\end{align}
In Fig.~\ref{fig:rabispread} we show how the ratio
$\Delta\Omega/\bar\Omega$ changes with sample temperature and misalignment
angle. For the temperatures and effective misalignment angles, we find
typical values of $\Delta\Omega/\bar\Omega$ in the range of 0.05-0.4.

An additional source of Rabi frequency inhomogeneity between different
lattice sites is the imperfect transmissivity of the lattice reflector
for the probe wavelength, introducing a standing wave component in the
probe beam. The wavelength mismatch between the lattice and probe
beams leads to a modulation of the residual probe standing wave
intensity between different lattice sites, and therefore different
Rabi frequencies. However, this effect is different from the
excitation induced inhomogeneity discussed above since it is
homogeneous within each site. The mirror reflectivity at the probe
wavelength was measured as $0.5\%$, leading to a
$\Delta\Omega/\bar\Omega$ contribution of 0.07. In the following
section, we will use the inhomogeneity information obtained from
spectroscopy to estimate inhomogeneity-induced clock frequency shifts
related to in-site Rabi frequency variation. The Rabi flopping data
thus overestimate the in-site Rabi frequency variation slightly.

\section{Inhomogeneity-induced Density Shifts}
\label{sec:density}

For $^{87}$Sr at $\mu$K temperatures, interatomic collisions are
suppressed due to Fermi statistics, which exclude collisions between
identical fermions resulting from even partial waves. However, using
$^{87}$Sr a non-zero density shift to the transition frequency was
recently measured~\cite{Campbell09}. At ultracold temperatures
$p$-wave and higher order odd partial wave collisions are frozen out,
and any remaining $s$-wave interactions should be suppressed. As
discussed in the preceding sections, the trapping potential will
result in an inhomogeneous excitation process unless all atoms are in
the same motional quantum state. Since they are fermions that have
been prepared in the same electronic and nuclear states, this is in
principle impossible and there will always be an inhomogeneous
excitation unless a lattice with at most one particle per site is
realized. The inhomogeneous excitation changes the quantum statistics
of the trapped atomic sample. This change in statistics leads to
distinguishable fermions even for $\mu$K temperatures and gives rise
to $s$-wave collisions, and therefore a density shift. However, we
will show that these small density-related frequency shifts can be
minimized by choosing an average excitation fraction close to 50\%.

The mean field energy shifts in the two state system consisting of
ground and excited clock state $\Ket{g}$ and $\Ket{e}$ can be expressed
as~\cite{Harber02}
\begin{align}
  \Delta E_g &= \frac{4\pi\hbar^2}{m}(G^{(2)}_{eg}a_{eg}\rho_{e} +
  G^{(2)}_{gg}a_{gg}\rho_{g})\\
  \Delta E_e &= \frac{4\pi\hbar^2}{m}(G^{(2)}_{ee}a_{ee}\rho_{e} +
  G^{(2)}_{eg}a_{eg}\rho_{g}),
\end{align}
where $G^{(2)}_{ij}$ ($a_{ij}$) is the two-body correlation function
(scattering length) between states $i$ and $j$, and $\rho_j$ is the
partial density of state $j$ in a trap site. The above expression is
applicable when the de Broglie wavelength of the scattering particles
is large compared to their scattering length.

The clock frequency shift can be calculated from the differential
energy shift $\Delta E_e-\Delta E_g$. For identical fermions
$G^{(2)}_{ii}$ vanishes. Its bosonic equivalent equals two and for
degenerate bosons it becomes unity. The clock frequency shift
$\Delta\nu$ in the two-state fermionic system is given by
\begin{equation}
\label{densityshiftsimple}
  h\Delta\nu = \frac{4\pi\hbar^2}{m} a_{eg} G^{(2)}_{eg} [\rho_g - \rho_e].
\end{equation}

We can estimate the magnitude of this shift by considering the
evolution of two representative atoms during the Rabi pulse. These two
atoms will follow slightly different trajectories on the Bloch sphere
since their Rabi frequencies differ through the inhomogeneities derived
in the previous sections, causing the energy difference between the
clock states to change during the clock pulse.

Under the clock pulse, the electronic state of particle $j$ evolves with the
time-dependent Hamiltonian
\begin{equation}
  \label{eq:29}
  H_j(t) = -\frac{h}{2} \{\Omega_j \sigma_x + [\delta-\Delta\nu(t)]\sigma_z\},
\end{equation}
where $\sigma_x$ ($\sigma_z$) is the first (third) Pauli matrix,
$\Omega_j$ is the Rabi frequency for particle $j$, and $\delta$ is the
probe laser detuning from the bare atomic resonance. We assume that
all particles are prepared in the excited electronic state $\Ket{e}$,
and use a discretized version of the Schr\"odinger equation to
propagate the wave function coefficients over a small time step
$\Delta t$:
\begin{equation}
  \label{eq:30}
  \Ket{\psi_j(t+\Delta t)} = \{I_2 - i \Delta t H_j(t)/\hbar\} \Ket{\psi_j(t)},
\end{equation}
where $I_2$ is the $2\times 2$ identity matrix. After each time step
$\Delta t$, the density shift $\Delta\nu(t)$ is recalculated for the
updated wave functions $\Ket{\psi_j(t)}\equiv \alpha_j(t)\Ket{e}+
\beta_j(t)\Ket{g}$ ($j=1,2$), according to
Eqn.~\ref{densityshiftsimple}. The antisymmetrized two-body
correlation function between the two-level fermions 1 and 2 is~\cite{Campbell09}
\begin{equation}
  \label{eq:24}
  G^{(2)}_{eg} = 1-|\alpha_1(t)\alpha^*_2(t)+\beta_1(t)\beta_2^*(t)|^2.
\end{equation}
A more detailed many-body model relating $G^{(2)}$ to transitions to
motional singlet and triplet states in the wave function is presented
in reference~\cite{Rey09}.

If we consider a ``mean particle'' (particle 1) evolving with Rabi
frequency $\bar\Omega$ and a typical perturbing particle (particle
2) evolving with $\bar\Omega + \Delta\Omega$, we can define an
approximate associated density shift as
\begin{equation}
  \label{eq:24}
  \Delta\nu(t) = \Delta\nu_0
  [1-|\alpha_1\alpha_2^*+\beta_1\beta_2^*|^2] [1-2|\alpha_1|^2],
\end{equation}
where $\Delta\nu_0 \equiv 2\hbar \rho_0 a_{eg} / m$, with average atom
density $\rho_0$. Here, the two-body correlation function is
multiplied by the inversion of the mean particle following a
trajectory on the Bloch sphere. Due to the background interaction with
the other particles, a mean field energy builds up during the pulse,
causing the detuning and thus the effective Rabi frequency to change
dynamically. Many collisions occur during the Rabi time since the rate
of collision attempts is given by the inverse of the transverse trap
frequency $1/\nu_r \simeq\unit{2}{ms}$ and the typical Rabi time for
clock operation is $\unit{80}{ms}$. This separation of time scales
makes the mean-field treatment applicable. For pulse times $\ll
1/\nu_r$, the above description of the excitation process would have
to be modified to include wave packets of vibrational states. For
these short pulse times, no collisions can occur which can be
explained by a local picture. The atoms are effectively confined to a
volume determined by their velocity and the pulse length. If this
volume is significantly smaller than the trap size, the local
excitation process is very homogeneous. No collisions can occur since
the atoms cannot travel far enough to encounter collision partners
that have been excited in a slightly different way.

\begin{figure}[htbp]
\centering
\includegraphics[width=\columnwidth]{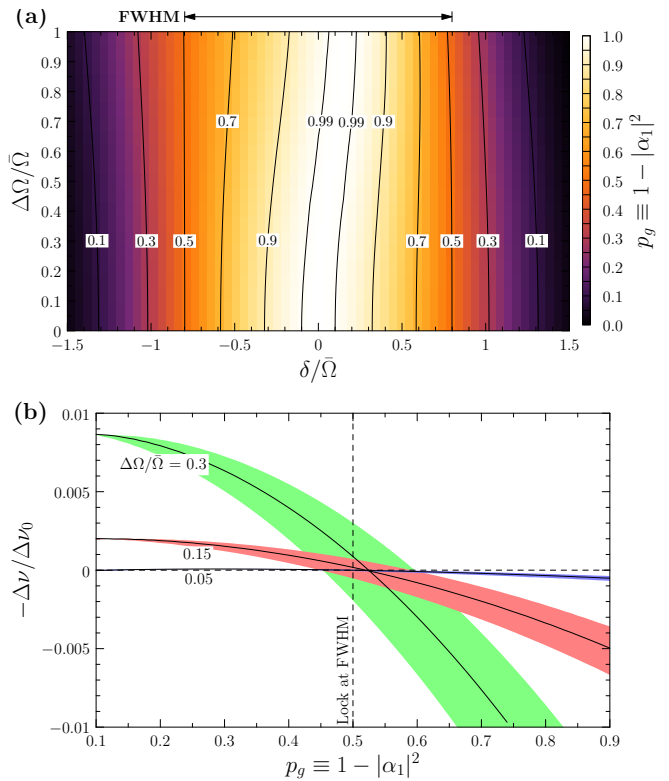}
\caption{(color online). (a) Contour plot of the ground state fraction
  $p_g = 1 - |\alpha_1|^2$ for the mean particle as a function of
  detuning in units of $\bar\Omega$ and inhomogeneity parameter
  $\Delta\Omega/\bar\Omega$ after a $\pi$-pulse. The solid curves are
  contours of equal $p_g$, visualizing the locking points of the
  spectroscopy laser. Note that the contours for locking close to the
  Full Width Half Maximum (FWHM) are insensitive to the inhomogeneity
  parameter. (b) The solid lines show the clock frequency shift (or
  locking error) with respect to the bare atomic transition in units
  of $\Delta\nu_0 = 2\hbar a_{eg} \rho_0/m$ as a function of final
  ground state fraction $p_g = 1 - |\alpha_1|^2$ after the pulse for
  the relevant range of $\Delta\Omega/\bar\Omega$. A negative sign was
  included in the shift to recreate the experimental behavior for a
  negative scattering length. The shift crosses zero close to $p_g =
  0.5$ and varies quadratically both with inhomogeneity parameter and
  ground state fraction. The shaded areas indicate the change in the
  solid curve when the $\pi$-pulse condition is allowed to vary by
  5\%.}
\label{fig:dshift}
\end{figure}

The clock transition frequency is measured by locking the spectroscopy
laser to points of equal height on the transition lineshape. To model
the experimental procedure, we include the mean-field density shift as
a time-dependent detuning in the wave function coefficients via
Eqn.~\ref{eq:30}. We calculate the line shape for a pulse time
corresponding to a $\pi$-pulse (on resonance for the mean particle) as
a function of the inhomogeneity parameter $\Delta\Omega/\bar\Omega$,
as shown in Fig.~\ref{fig:dshift}(a). The measured clock shift can be
visualized by comparing two contour lines of equal excited state
fraction $p_e$. The spectroscopy laser is locked to their average
detuning which changes with both the inhomogeneity and the pair of
contours chosen. The final frequency offset is negative for $p_g<0.52$,
vanishes at $p_g \simeq 0.52$ and becomes positive for $p_g>0.52$. The
resulting frequency shift (or locking error) with respect to the bare
atomic transition in units of $\Delta\nu_0$ is shown in
Fig.~\ref{fig:dshift}(b) for the experimentally relevant range of
$\Delta\Omega/\bar\Omega$. The frequency shift is shown with a
negative sign to recreate the behavior for a negative scattering
length $a_{eg}$.

In using Eqn.~\ref{eq:24} and detecting only the mean particle
excitation fraction, we have chosen a maximally asymmetric model where
the representative particle is only perturbed by the background
particles but does not act back on the perturbing particles. A
maximally symmetric model can be derived by choosing representative
Rabi frequencies $\bar\Omega\pm\Delta\Omega$, and using the mean of
both excited state fractions in Eqn.~\ref{eq:24} as well as in the
detection of the line shape. In this two-particle model, there is no
preferred particle and the resulting lineshape exhibits loss of
contrast as the inhomogeneity grows. We find that the density shift
calculated from such lineshapes shows a very similar dependence
on the ground state fraction and its magnitude is the same.

To extract information about $a_{eg}$ from experimental data for small
inhomogeneity, we fit the numerical clock shift data with a with a
simple polynomial. The frequency shift increases quadratically with
$\Delta\Omega/\bar\Omega$, is quadratic in $p_g$ and changes sign
close to $p_g=0.5$. Fitting the functional form
\begin{equation}
  \label{eq:10}
  A (p_g-B)(p_g-C) \left(\Delta\Omega/\bar\Omega\right)^2
\end{equation}
to numerical data obtained after a $\pi$-pulse excitation, as in
Fig.~\ref{fig:dshift}(b) for $\Delta\Omega/\bar\Omega \in [0,0.4]$,
results in $A = -0.410(3)$, $B = 0.528(1)$, and $C = -0.443(7)$ with RMS
error $3\times10^{-4}$. To a good approximation, we can thus use the
formula
\begin{equation}
  \label{eq:20}
  \Delta\nu_\pi \simeq - \Delta\nu_0
  \left(\Delta\Omega/\bar\Omega\right)^2 0.41 (p_g - 0.53)(p_g + 0.44)
\end{equation}
to calculate the density shift for small inhomogeneity, measured by
locking the spectroscopy laser to points of equal ground state
fraction after a $\pi$-pulse (defined on resonance for the mean
particle) for particles initially prepared in the excited electronic
state. However, if the $\pi$-pulse condition is allowed to vary by
only 5\%, the density shift curve changes as indicated by the shaded
areas in Fig~\ref{fig:dshift}(b). Since we cannot exclude such
variation over the course of a day, the zero crossing is not well
defined and obtains an uncertainty according to how well the probe
intensity is stabilized on long timescales. In the following, we will
use the standard error given by fitting Eqn.~\ref{eq:20} to
experimental data using weighted least-squares to indicate
experimental uncertainty and add an estimated relative amplitude error
of 30\% in quadrature to account for the $\pi$-pulse condition
uncertainty.

\begin{figure}[htpb]
  \centering
  \includegraphics[width=\columnwidth]{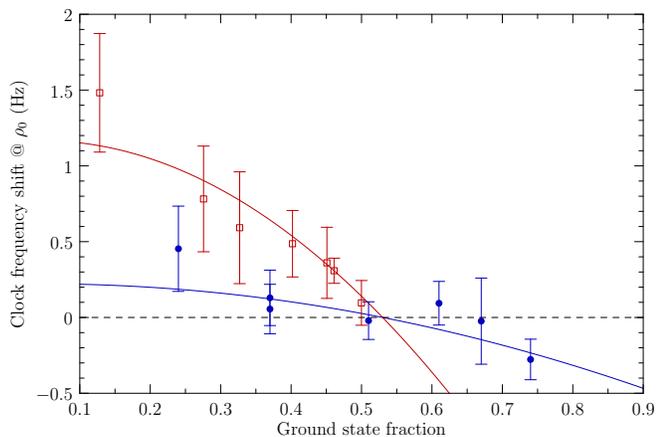}
  \caption{(color online). Clock frequency shift for \unit{1}{\mu K}
    (circles) and \unit{3}{\mu K} (open squares), normalized to the
    typical operating density $\rho_0 = \unit{1\times
      10^{11}}{cm^{-3}}$. The solid lines are fits of Eqn.~\ref{eq:20}
    to the data with the amplitude as the only fit parameter.}
  \label{fig:dshift-data}
\end{figure}

The density shift data from Campbell~\textit{et al.}~\cite{Campbell09}
is shown in Fig.~\ref{fig:dshift-data} along with fits based on
Eqn.~\ref{eq:20}, adjusting the amplitude of the polynomial in $p_g$
as the only fit parameter. For temperatures $T_1=\unit{1}{\mu K}$ and
$T_2=\unit{3}{\mu K}$, we find fit amplitudes of
$\Delta\nu(T_1)=\unit{-2.3\pm 0.7}{Hz}$ and
$\Delta\nu(T_2)=\unit{-12.1\pm 0.9}{Hz}$, respectively. Although we
can exclude $p$-wave scattering contributions~\cite{Campbell09}, the
corresponding values of $a_{eg}$ are larger than the unitarity limit
given by $a_{eg} = \lambda_T/2\pi$, with thermal wavelength $\lambda_T
= h/\sqrt{2\pi m k_B (T + T_\text{zp})}$ and zero-point temperature
$T_\text{zp}\simeq\unit{3.5}{\mu K}$ corresponding to the ground state
energy of the trapping potential. On the other hand, in the unitarity
limit, the ratio of both shift coefficients scales as
\begin{equation}
  \label{eq:21}
  \left[\frac{(\Delta\Omega/\bar\Omega)_1}{(\Delta\Omega/\bar\Omega)_2}\right]^2
  \sqrt{\frac{T_2+T_\text{zp}}{T_1+T_\text{zp}}} \simeq 0.14,
\end{equation}
which agrees with the ratio of measured shift coefficients
$\Delta\nu(T_1)/\Delta\nu(T_2) = 0.2(1)$ for
$(\Delta\Omega/\bar\Omega)_1=0.05$ and
$(\Delta\Omega/\bar\Omega)_2=0.14$ at $\Delta\theta =
\unit{10}{mrad}$.

The current two-particle model describes the qualitative behavior of
the experimental data well and the scaling factors derived agree with
the unitarity limit, but cannot explain the large density shift
quantitatively. This discrepancy probably stems from applying
Eqn.~\ref{densityshiftsimple} in a dynamic context. Even though there
are many collisions during a typical Rabi pulse time,
Eqn.~\ref{densityshiftsimple} describes the mean field differential
energy shift in a two-level system in steady state and cannot model
the lineshape evolution during the Rabi pulse accurately. A full
many-body model is expected to improve our understanding of the
density shift considerably~\cite{Rey09}. Even with a better model,
extracting reliable information about the scattering lengths from
density shift data remains a challenging problem.

\section{Conclusion}
\label{sec:conclusion}

In this paper we have investigated the inhomogeneous excitation
introduced by the transverse degrees of freedom for Rabi spectroscopy
in a 1D optical lattice. An analytical model for how the longitudinal
sidebands are influenced by the finite temperature of the atomic
sample is developed, and is used to allow us to extract temperatures
from sideband scans. Furthermore, by modeling and fitting Rabi
oscillations and lineshapes the degree of inhomogeneity introduced by
transverse temperature and an effective probe beam misalignment can be
measured. The inhomogeneity directly affects the spectroscopic process
and causes otherwise identical ultracold fermions to become
distinguishable during the excitation process. This effect is modelled
by a time dependent two-particle correlation function giving rise to
density-dependent clock frequency shifts. By determining the
temperature and inhomogeneity of the atomic sample as described here,
and by controlling the excitation fraction, this density shift can now
be measured, controlled and zeroed. We have presented a simple
two-particle model that models the density shift data qualitatively.

We acknowledge early contributions by T. Zelevinsky to the discussion
on probe inhomogeneities, and technical contributions by T. L.
Nicholson and M. H. G. de Miranda. We thank A. M. Rey and K. Gibble
for insightful discussions. This work was supported by funding from
NIST, NSF, DARPA, and ONR.

\appendix

\section{Sideband Lineshape Derivation}
\label{sec:lineshape-deriv}

The derivation in this appendix generalizes arguments presented by
Boyd~\cite{Boyd07b}, Le Targat~\cite{Letargat07}, and
Ludlow~\cite{Ludlow08b} to include longitudinal anharmonicity and
population in higher bands.

An atom's radial motional state $(n_x,n_y)$ shifts the sideband
transition frequency via the $r^2z^2$ coupling term in the site
potential. For a fixed longitudinal quantum number $n_z$ and detuning
$\delta$, only atoms with specific longitudinal quantum numbers
$(n_x,n_y)$ will be resonant with the probe laser. This resonance
condition is given by the longitudinal energy gap $\gamma$
(Eqn.~\ref{eq:15}). The sideband amplitude at this detuning is
determined by the Boltzmann factors for the corresponding quantum
numbers. The single-$n_z$ sideband lineshape is a superposition of
individual atomic transitions between states $\Ket{i} = \Ket{n_x,n_y}$
and $\Ket{f}=\Ket{n_x,n_y,n_z+1}$, weighted by the Boltzmann
distribution. The individual atomic transition is power broadened with
base width $\Gamma$, given by the carrier Rabi frequency, since the
natural transition linewidth is negligible. We also neglect
higher-order contributions to the sideband where the change in
longitudinal quantum number is accompanied by a change in the radial
quantum numbers since their amplitude is suppressed by an additional
factor of $\eta_x^2$. We find
\begin{equation}
  \label{eq:9}
  \sigma^{n_z}_\text{blue}(\delta) \propto \sum_{n_x,n_y}
  \frac{{z_r}^{n_x}{z_r}^{n_y}}{1+\frac{4}{\Gamma^2}[\delta-\gamma(n_z)]^2},
\end{equation}
where $z_r = \exp[-h\nu_r/(k_B T_r)]$ is the Boltzmann factor
associated with transverse state $n_x$ ($n_y$) and transverse
temperature $T_r$. By introducing a radial quantum number $n_r \equiv
n_x + n_y$ and corresponding two-dimensional degeneracy factor
$(n_r+1)$, we can reduce the summation to one dimension:
\begin{equation}
  \label{eq:26}
  \sigma^{n_z}_\text{blue}(\delta) \propto \sum_{n_r} \frac{(n_r+1)
    {z_r}^{n_r+1}}{1+\frac{4}{\Gamma^2}[\delta-\tilde\gamma(n_z)+\nu_\text{rec}
    \frac{\nu_r}{\nu_z}(n_r+1)]^2},
\end{equation}
with base longitudinal gap $\tilde\gamma(n_z) \equiv \nu_z
-\nu_\text{rec}(n_z+1)$.

For detunings $\delta/\gamma(n_z)\ll 1$ on the shallow sideband
slope facing the carrier, the sum's main term will be associated with
the quantum number $n_r$ that minimizes the denominator, leading to a
relation between detuning and quantum number:
\begin{equation}
  \label{eq:25}
  n_r + 1 \simeq -\frac{\nu_z}{\nu_\text{rec}\nu_r}
  [\delta-\tilde\gamma(n_z)].
\end{equation}
We approximate the sum by this main term and find
\begin{equation}
  \label{eq:27}
  \sigma^{n_z}_\text{blue}(\delta) \propto
  \left[1-\frac{\delta}{\tilde\gamma(n_z)}\right]
  e^{-\alpha[1-\delta/\tilde\gamma(n_z)]}
  \Theta[\tilde\gamma(n_z)-\delta],
\end{equation}
with
$\alpha\equiv\frac{\tilde\gamma(n_z)}{\nu_\text{rec}}\frac{h\nu_z}{k_BT_r}$.
We ensure the applicability of the approximation by cutting off the
lineshape at $\delta=\tilde\gamma(n_z)$ with the Heaviside function
$\Theta$. For the relevant parameter ranges, the lineshape can be
approximately area-normalized to three significant figures with a
prefactor $\alpha^2/\tilde\gamma(n_z)$.

The longitudinal trap anharmonicity places each of these single-band
component lineshapes at slightly different detunings. We find the
final blue sideband lineshape as a Boltzmann-weighted superposition of
area-normalized components:
\begin{equation}
  \label{eq:28}
  \begin{split}
  \sigma_\text{blue}(\delta) &=
  \left(\sum_{n_z}^{N_z}e^{-\frac{E_{n_z}}{k_B T_z}}\right)^{-1}
  \sum_{n_z}^{N_z}
  e^{-\frac{E_{n_z}}{k_B T_z}} \sigma_\text{blue}^{n_z}(\delta)\\
  E_{n_z}/h &= \nu_z(n_z+\frac{1}{2}) - \frac{\nu_\text{rec}}{2}(n_z^2+n_z+1).
  \end{split}
\end{equation}
Here, the quartic approximation to the longitudinal potential
reproduces the lowest energies and gaps to within a few percent. A
better approximation to the energy $E_{n}$ of the $n$-th longitudinal
state is given by $E_{n}/(h\nu_\text{rec}) = [a_n(q)+b_{n+1}(q)]/2
+2q$, where $a_n$ and $b_{n+1}$ are the characteristic values of the
Mathieu equation bounding its $n$-th stability region at parameter
value $q$ (see Ref.~\cite{Slater52}, for
example). Here $q$ has to be optimized numerically -- starting from
the harmonic approximation $q_0=U_0/(4h\nu_\text{rec})$ -- such that
the lowest gap $(E_1-E_0)/h \simeq \nu_\text{rec}[a_1(q)-a_0(q)]$ is
equal to the measured longitudinal trap frequency.


\begin{thebibliography}{27}
\expandafter\ifx\csname natexlab\endcsname\relax\def\natexlab#1{#1}\fi
\expandafter\ifx\csname bibnamefont\endcsname\relax
\def\bibnamefont#1{#1}\fi
\expandafter\ifx\csname bibfnamefont\endcsname\relax
  \def\bibfnamefont#1{#1}\fi
\expandafter\ifx\csname citenamefont\endcsname\relax
  \def\citenamefont#1{#1}\fi
\expandafter\ifx\csname url\endcsname\relax
  \def\url#1{\texttt{#1}}\fi
\expandafter\ifx\csname urlprefix\endcsname\relax\def\urlprefix{URL }\fi
\providecommand{\bibinfo}[2]{#2}
\providecommand{\eprint}[2][]{\url{#2}}

\bibitem[{\citenamefont{Ludlow et~al.}(2008)\citenamefont{Ludlow, Zelevinsky,
  Campbell, Blatt, Boyd, de~Miranda, Martin, Thomsen, Foreman, Ye
  et~al.}}]{Ludlow08}
\bibinfo{author}{\bibfnamefont{A.~D.} \bibnamefont{Ludlow}},
  \bibinfo{author}{\bibfnamefont{T.}~\bibnamefont{Zelevinsky}},
  \bibinfo{author}{\bibfnamefont{G.~K.} \bibnamefont{Campbell}},
  \bibinfo{author}{\bibfnamefont{S.}~\bibnamefont{Blatt}},
  \bibinfo{author}{\bibfnamefont{M.~M.} \bibnamefont{Boyd}},
  \bibinfo{author}{\bibfnamefont{M.~H.~G.} \bibnamefont{de~Miranda}},
  \bibinfo{author}{\bibfnamefont{M.~J.} \bibnamefont{Martin}},
  \bibinfo{author}{\bibfnamefont{J.~W.} \bibnamefont{Thomsen}},
  \bibinfo{author}{\bibfnamefont{S.~M.} \bibnamefont{Foreman}},
  \bibinfo{author}{\bibfnamefont{J.}~\bibnamefont{Ye}}, \bibnamefont{et~al.},
  \bibinfo{journal}{Science} \textbf{\bibinfo{volume}{319}},
  \bibinfo{pages}{1805} (\bibinfo{year}{2008}).

\bibitem[{\citenamefont{Campbell et~al.}(2008)\citenamefont{Campbell, Ludlow,
  Blatt, Thomsen, Martin, de~Miranda, Zelevinsky, Boyd, Ye, Diddams
  et~al.}}]{Campbell08}
\bibinfo{author}{\bibfnamefont{G.~K.} \bibnamefont{Campbell}},
  \bibinfo{author}{\bibfnamefont{A.~D.} \bibnamefont{Ludlow}},
  \bibinfo{author}{\bibfnamefont{S.}~\bibnamefont{Blatt}},
  \bibinfo{author}{\bibfnamefont{J.~W.} \bibnamefont{Thomsen}},
  \bibinfo{author}{\bibfnamefont{M.~J.} \bibnamefont{Martin}},
  \bibinfo{author}{\bibfnamefont{M.~H.~G.} \bibnamefont{de~Miranda}},
  \bibinfo{author}{\bibfnamefont{T.}~\bibnamefont{Zelevinsky}},
  \bibinfo{author}{\bibfnamefont{M.~M.} \bibnamefont{Boyd}},
  \bibinfo{author}{\bibfnamefont{J.}~\bibnamefont{Ye}},
  \bibinfo{author}{\bibfnamefont{S.~A.} \bibnamefont{Diddams}},
  \bibnamefont{et~al.}, \bibinfo{journal}{Metrologia}
  \textbf{\bibinfo{volume}{45}}, \bibinfo{pages}{539} (\bibinfo{year}{2008}).

\bibitem[{\citenamefont{Heavner et~al.}(2005)\citenamefont{Heavner, Jefferts,
  Donley, Shirley, and Parker}}]{Heavner05}
\bibinfo{author}{\bibfnamefont{T.~P.} \bibnamefont{Heavner}},
  \bibinfo{author}{\bibfnamefont{S.~R.} \bibnamefont{Jefferts}},
  \bibinfo{author}{\bibfnamefont{E.~A.} \bibnamefont{Donley}},
  \bibinfo{author}{\bibfnamefont{J.~H.} \bibnamefont{Shirley}},
  \bibnamefont{and} \bibinfo{author}{\bibfnamefont{T.~E.}
  \bibnamefont{Parker}}, \bibinfo{journal}{Metrologia}
  \textbf{\bibinfo{volume}{42}}, \bibinfo{pages}{411} (\bibinfo{year}{2005}).

\bibitem[{\citenamefont{Bize et~al.}(2005)\citenamefont{Bize, Laurent, Abgrall,
  Marion, Maksimovic, Cacciapuoti, Gr\"{u}nert, Vian, {Pereira Dos Santos},
  Rosenbusch et~al.}}]{Bize05}
\bibinfo{author}{\bibfnamefont{S.}~\bibnamefont{Bize}},
  \bibinfo{author}{\bibfnamefont{P.}~\bibnamefont{Laurent}},
  \bibinfo{author}{\bibfnamefont{M.}~\bibnamefont{Abgrall}},
  \bibinfo{author}{\bibfnamefont{H.}~\bibnamefont{Marion}},
  \bibinfo{author}{\bibfnamefont{I.}~\bibnamefont{Maksimovic}},
  \bibinfo{author}{\bibfnamefont{L.}~\bibnamefont{Cacciapuoti}},
  \bibinfo{author}{\bibfnamefont{J.}~\bibnamefont{Gr\"{u}nert}},
  \bibinfo{author}{\bibfnamefont{C.}~\bibnamefont{Vian}},
  \bibinfo{author}{\bibfnamefont{F.}~\bibnamefont{{Pereira Dos Santos}}},
  \bibinfo{author}{\bibfnamefont{P.}~\bibnamefont{Rosenbusch}},
  \bibnamefont{et~al.}, \bibinfo{journal}{J. Phys. B}
  \textbf{\bibinfo{volume}{38}}, \bibinfo{pages}{S449} (\bibinfo{year}{2005}).

\bibitem[{\citenamefont{Weyers et~al.}(2001)\citenamefont{Weyers, H\"{u}bner,
  Schr\"{o}der, Tamm, and Bauch}}]{Weyers01}
\bibinfo{author}{\bibfnamefont{S.}~\bibnamefont{Weyers}},
  \bibinfo{author}{\bibfnamefont{U.}~\bibnamefont{H\"{u}bner}},
  \bibinfo{author}{\bibfnamefont{R.}~\bibnamefont{Schr\"{o}der}},
  \bibinfo{author}{\bibfnamefont{C.}~\bibnamefont{Tamm}}, \bibnamefont{and}
  \bibinfo{author}{\bibfnamefont{A.}~\bibnamefont{Bauch}},
  \bibinfo{journal}{Metrologia} \textbf{\bibinfo{volume}{38}},
  \bibinfo{pages}{343} (\bibinfo{year}{2001}).

\bibitem[{\citenamefont{Boyd et~al.}(2006)\citenamefont{Boyd, Zelevinsky,
  Ludlow, Foreman, Blatt, Ido, and Ye}}]{Boyd06}
\bibinfo{author}{\bibfnamefont{M.~M.} \bibnamefont{Boyd}},
  \bibinfo{author}{\bibfnamefont{T.}~\bibnamefont{Zelevinsky}},
  \bibinfo{author}{\bibfnamefont{A.~D.} \bibnamefont{Ludlow}},
  \bibinfo{author}{\bibfnamefont{S.~M.} \bibnamefont{Foreman}},
  \bibinfo{author}{\bibfnamefont{S.}~\bibnamefont{Blatt}},
  \bibinfo{author}{\bibfnamefont{T.}~\bibnamefont{Ido}}, \bibnamefont{and}
  \bibinfo{author}{\bibfnamefont{J.}~\bibnamefont{Ye}},
  \bibinfo{journal}{Science} \textbf{\bibinfo{volume}{314}},
  \bibinfo{pages}{1430} (\bibinfo{year}{2006}).

\bibitem[{\citenamefont{Ye et~al.}(2008)\citenamefont{Ye, Kimble, and
  Katori}}]{Ye08}
\bibinfo{author}{\bibfnamefont{J.}~\bibnamefont{Ye}},
  \bibinfo{author}{\bibfnamefont{H.~J.} \bibnamefont{Kimble}},
  \bibnamefont{and} \bibinfo{author}{\bibfnamefont{H.}~\bibnamefont{Katori}},
  \bibinfo{journal}{Science} \textbf{\bibinfo{volume}{320}},
  \bibinfo{pages}{1734} (\bibinfo{year}{2008}).

\bibitem[{\citenamefont{Wineland and Itano}(1979)}]{Wineland79}
\bibinfo{author}{\bibfnamefont{D.~J.} \bibnamefont{Wineland}} \bibnamefont{and}
  \bibinfo{author}{\bibfnamefont{W.~M.} \bibnamefont{Itano}},
  \bibinfo{journal}{Phys. Rev. A} \textbf{\bibinfo{volume}{20}},
  \bibinfo{pages}{1521} (\bibinfo{year}{1979}).

\bibitem[{\citenamefont{Leibfried et~al.}(2003)\citenamefont{Leibfried, Blatt,
  Monroe, and Wineland}}]{Leibfried03}
\bibinfo{author}{\bibfnamefont{D.}~\bibnamefont{Leibfried}},
  \bibinfo{author}{\bibfnamefont{R.}~\bibnamefont{Blatt}},
  \bibinfo{author}{\bibfnamefont{C.}~\bibnamefont{Monroe}}, \bibnamefont{and}
  \bibinfo{author}{\bibfnamefont{D.}~\bibnamefont{Wineland}},
  \bibinfo{journal}{Rev. Mod. Phys.} \textbf{\bibinfo{volume}{75}},
  \bibinfo{pages}{281} (\bibinfo{year}{2003}).

\bibitem[{\citenamefont{Campbell et~al.}(2009)\citenamefont{Campbell, Boyd,
  Thomsen, Martin, Blatt, Swallows, Nicholson, Fortier, Oates, Diddams
  et~al.}}]{Campbell09}
\bibinfo{author}{\bibfnamefont{G.~K.} \bibnamefont{Campbell}},
  \bibinfo{author}{\bibfnamefont{M.~M.} \bibnamefont{Boyd}},
  \bibinfo{author}{\bibfnamefont{J.~W.} \bibnamefont{Thomsen}},
  \bibinfo{author}{\bibfnamefont{M.~J.} \bibnamefont{Martin}},
  \bibinfo{author}{\bibfnamefont{S.}~\bibnamefont{Blatt}},
  \bibinfo{author}{\bibfnamefont{M.~D.} \bibnamefont{Swallows}},
  \bibinfo{author}{\bibfnamefont{T.~L.} \bibnamefont{Nicholson}},
  \bibinfo{author}{\bibfnamefont{T.}~\bibnamefont{Fortier}},
  \bibinfo{author}{\bibfnamefont{C.~W.} \bibnamefont{Oates}},
  \bibinfo{author}{\bibfnamefont{S.~A.} \bibnamefont{Diddams}},
  \bibnamefont{et~al.}, \bibinfo{journal}{Science}
  \textbf{\bibinfo{volume}{324}}, \bibinfo{pages}{360} (\bibinfo{year}{2009}).

\bibitem[{\citenamefont{{A. M. Rey, A. V. Gorshkov, and C. Rubbo}}(2009)}]{Rey09}
\bibinfo{author}{\bibnamefont{{A. M. Rey, A. V. Gorshkov, and C. Rubbo}}}, \bibinfo{journal}{
  arXiv:0907.2245v2}  (\bibinfo{year}{2009}).

\bibitem[{\citenamefont{{K. Gibble}}(2009)}]{Gibble09}
\bibinfo{author}{\bibnamefont{{K. Gibble}}}, \bibinfo{journal}{
  arXiv:0908.3147v1}  (\bibinfo{year}{2009}).


\bibitem[{\citenamefont{Baillard et~al.}(2008)\citenamefont{Baillard,
  Fouch{\'e}, {Le Targat}, Westergaard, Lecallier, Chapelet, Abgrall, Rovera,
  Laurent, Rosenbusch et~al.}}]{Baillard08}
\bibinfo{author}{\bibfnamefont{X.}~\bibnamefont{Baillard}},
  \bibinfo{author}{\bibfnamefont{M.}~\bibnamefont{Fouch{\'e}}},
  \bibinfo{author}{\bibfnamefont{R.}~\bibnamefont{{Le Targat}}},
  \bibinfo{author}{\bibfnamefont{P.~G.} \bibnamefont{Westergaard}},
  \bibinfo{author}{\bibfnamefont{A.}~\bibnamefont{Lecallier}},
  \bibinfo{author}{\bibfnamefont{F.}~\bibnamefont{Chapelet}},
  \bibinfo{author}{\bibfnamefont{M.}~\bibnamefont{Abgrall}},
  \bibinfo{author}{\bibfnamefont{G.~D.} \bibnamefont{Rovera}},
  \bibinfo{author}{\bibfnamefont{P.}~\bibnamefont{Laurent}},
  \bibinfo{author}{\bibfnamefont{P.}~\bibnamefont{Rosenbusch}},
  \bibnamefont{et~al.}, \bibinfo{journal}{Europhys. J. D}
  \textbf{\bibinfo{volume}{48}}, \bibinfo{pages}{11} (\bibinfo{year}{2008}).

\bibitem[{\citenamefont{Akatsuka et~al.}(2008)\citenamefont{Akatsuka, Takamoto,
  and Katori}}]{Akatsuka09}
\bibinfo{author}{\bibfnamefont{T.}~\bibnamefont{Akatsuka}},
  \bibinfo{author}{\bibfnamefont{M.}~\bibnamefont{Takamoto}}, \bibnamefont{and}
  \bibinfo{author}{\bibfnamefont{H.}~\bibnamefont{Katori}},
  \bibinfo{journal}{Nature Physics} \textbf{\bibinfo{volume}{4}},
  \bibinfo{pages}{954} (\bibinfo{year}{2008}).

\bibitem[{\citenamefont{Boyd}(2007)}]{Boyd07b}
\bibinfo{author}{\bibfnamefont{M.~M.} \bibnamefont{Boyd}}, Ph.D. thesis,
  \bibinfo{school}{University of Colorado} (\bibinfo{year}{2007}),
  \urlprefix\url{http://jilawww.colorado.edu/yelabs/pubs/theses.html}.

\bibitem[{\citenamefont{Ludlow}(2008)}]{Ludlow08b}
\bibinfo{author}{\bibfnamefont{A.~D.} \bibnamefont{Ludlow}}, Ph.D. thesis,
  \bibinfo{school}{University of Colorado} (\bibinfo{year}{2008}),
  \urlprefix\url{http://jilawww.colorado.edu/yelabs/pubs/theses.html}.

\bibitem[{\citenamefont{Mukaiyama et~al.}(2003)\citenamefont{Mukaiyama, Katori,
  Ido, Li, and Kuwata-Gonokami}}]{Mukaiyama03}
\bibinfo{author}{\bibfnamefont{T.}~\bibnamefont{Mukaiyama}},
  \bibinfo{author}{\bibfnamefont{H.}~\bibnamefont{Katori}},
  \bibinfo{author}{\bibfnamefont{T.}~\bibnamefont{Ido}},
  \bibinfo{author}{\bibfnamefont{Y.}~\bibnamefont{Li}}, \bibnamefont{and}
  \bibinfo{author}{\bibfnamefont{M.}~\bibnamefont{Kuwata-Gonokami}},
  \bibinfo{journal}{Phys. Rev. Lett.} \textbf{\bibinfo{volume}{90}},
  \bibinfo{pages}{113002} (\bibinfo{year}{2003}).

\bibitem[{\citenamefont{Loftus et~al.}(2004)\citenamefont{Loftus, Ido, Ludlow,
  Boyd, and Ye}}]{Loftus04}
\bibinfo{author}{\bibfnamefont{T.~H.} \bibnamefont{Loftus}},
  \bibinfo{author}{\bibfnamefont{T.}~\bibnamefont{Ido}},
  \bibinfo{author}{\bibfnamefont{A.~D.} \bibnamefont{Ludlow}},
  \bibinfo{author}{\bibfnamefont{M.~M.} \bibnamefont{Boyd}}, \bibnamefont{and}
  \bibinfo{author}{\bibfnamefont{J.}~\bibnamefont{Ye}}, \bibinfo{journal}{Phys.
  Rev. Lett.} \textbf{\bibinfo{volume}{93}}, \bibinfo{pages}{073003}
  (\bibinfo{year}{2004}).

\bibitem[{\citenamefont{Ludlow et~al.}(2006)\citenamefont{Ludlow, Huang,
  Notcutt, Zanon-Willette, Foreman, Boyd, Blatt, and Ye}}]{Ludlow07}
\bibinfo{author}{\bibfnamefont{A.~D.} \bibnamefont{Ludlow}},
  \bibinfo{author}{\bibfnamefont{X.}~\bibnamefont{Huang}},
  \bibinfo{author}{\bibfnamefont{M.}~\bibnamefont{Notcutt}},
  \bibinfo{author}{\bibfnamefont{T.}~\bibnamefont{Zanon-Willette}},
  \bibinfo{author}{\bibfnamefont{S.~M.} \bibnamefont{Foreman}},
  \bibinfo{author}{\bibfnamefont{M.~M.} \bibnamefont{Boyd}},
  \bibinfo{author}{\bibfnamefont{S.}~\bibnamefont{Blatt}}, \bibnamefont{and}
  \bibinfo{author}{\bibfnamefont{J.}~\bibnamefont{Ye}}, \bibinfo{journal}{Opt.
  Lett.} \textbf{\bibinfo{volume}{32}}, \bibinfo{pages}{641}
  (\bibinfo{year}{2006}).

\bibitem[{\citenamefont{Boyd et~al.}(2007)\citenamefont{Boyd, Zelevinsky,
  Ludlow, Blatt, Zanon-Willette, Foreman, and Ye}}]{Boyd07}
\bibinfo{author}{\bibfnamefont{M.~M.} \bibnamefont{Boyd}},
  \bibinfo{author}{\bibfnamefont{T.}~\bibnamefont{Zelevinsky}},
  \bibinfo{author}{\bibfnamefont{A.~D.} \bibnamefont{Ludlow}},
  \bibinfo{author}{\bibfnamefont{S.}~\bibnamefont{Blatt}},
  \bibinfo{author}{\bibfnamefont{T.}~\bibnamefont{Zanon-Willette}},
  \bibinfo{author}{\bibfnamefont{S.~M.} \bibnamefont{Foreman}},
  \bibnamefont{and} \bibinfo{author}{\bibfnamefont{J.}~\bibnamefont{Ye}},
  \bibinfo{journal}{Phys. Rev. A} \textbf{\bibinfo{volume}{76}},
  \bibinfo{pages}{022510} (\bibinfo{year}{2007}).

\bibitem[{\citenamefont{Porsev and Derevianko}(2004)}]{Porsev04}
\bibinfo{author}{\bibfnamefont{S.~G.} \bibnamefont{Porsev}} \bibnamefont{and}
  \bibinfo{author}{\bibfnamefont{A.}~\bibnamefont{Derevianko}},
  \bibinfo{journal}{Phys. Rev. A} \textbf{\bibinfo{volume}{69}},
  \bibinfo{pages}{042506} (\bibinfo{year}{2004}).

\bibitem[{\citenamefont{Santra et~al.}(2004)\citenamefont{Santra, Christ, and
  Greene}}]{Santra04}
\bibinfo{author}{\bibfnamefont{R.}~\bibnamefont{Santra}},
  \bibinfo{author}{\bibfnamefont{K.~V.} \bibnamefont{Christ}},
  \bibnamefont{and} \bibinfo{author}{\bibfnamefont{C.~H.}
  \bibnamefont{Greene}}, \bibinfo{journal}{Phys. Rev. A}
  \textbf{\bibinfo{volume}{69}}, \bibinfo{pages}{042510}
  (\bibinfo{year}{2004}).

\bibitem[{\citenamefont{Lemonde and Wolf}(2005)}]{Lemonde03}
\bibinfo{author}{\bibfnamefont{P.}~\bibnamefont{Lemonde}} \bibnamefont{and}
  \bibinfo{author}{\bibfnamefont{P.}~\bibnamefont{Wolf}},
  \bibinfo{journal}{Phys. Rev. A} \textbf{\bibinfo{volume}{72}},
  \bibinfo{pages}{033409} (\bibinfo{year}{2005}).

\bibitem[{\citenamefont{Gl{\"u}ck et~al.}(2002)\citenamefont{Gl{\"u}ck,
  Kolovsky, and Korsch}}]{Glueck02}
\bibinfo{author}{\bibfnamefont{M.}~\bibnamefont{Gl{\"u}ck}},
  \bibinfo{author}{\bibfnamefont{A.~R.} \bibnamefont{Kolovsky}},
  \bibnamefont{and} \bibinfo{author}{\bibfnamefont{H.~J.}
  \bibnamefont{Korsch}}, \bibinfo{journal}{Phys. Rep.}
  \textbf{\bibinfo{volume}{366}}, \bibinfo{pages}{102} (\bibinfo{year}{2002}).

\bibitem[{\citenamefont{Ludlow et~al.}(2006)\citenamefont{Ludlow,
  Boyd, Zelevinsky, Foreman, Blatt, Notcutt, Ido, and Ye}}]{Ludlow06}
\bibinfo{author}{\bibfnamefont{A.~D.} \bibnamefont{Ludlow}},
  \bibinfo{author}{\bibfnamefont{M.~M.}~\bibnamefont{Boyd}},
  \bibinfo{author}{\bibfnamefont{T.}~\bibnamefont{Zelevinsky}},
  \bibinfo{author}{\bibfnamefont{S.~M.}~\bibnamefont{Foreman}},
  \bibinfo{author}{\bibfnamefont{S.}~\bibnamefont{Blatt}},
  \bibinfo{author}{\bibfnamefont{M.}~\bibnamefont{Notcutt}},
  \bibinfo{author}{\bibfnamefont{T.}~\bibnamefont{Ido}},
  \bibnamefont{and} \bibinfo{author}{\bibfnamefont{J.}~\bibnamefont{Ye}},
  \bibinfo{journal}{Phys. Rev. Lett.} \textbf{\bibinfo{volume}{96}},
  \bibinfo{pages}{033003} (\bibinfo{year}{2006}).

\bibitem[{\citenamefont{Harber et~al.}(2002)\citenamefont{Harber, Lewandowski,
  McGuirk, and Cornell}}]{Harber02}
\bibinfo{author}{\bibfnamefont{D.~M.} \bibnamefont{Harber}},
  \bibinfo{author}{\bibfnamefont{H.~J.} \bibnamefont{Lewandowski}},
  \bibinfo{author}{\bibfnamefont{J.~M.} \bibnamefont{McGuirk}},
  \bibnamefont{and} \bibinfo{author}{\bibfnamefont{E.~A.}
  \bibnamefont{Cornell}}, \bibinfo{journal}{Phys. Rev. A}
  \textbf{\bibinfo{volume}{66}}, \bibinfo{pages}{053616}
  (\bibinfo{year}{2002}).

\bibitem[{\citenamefont{{Le Targat}}(2007)}]{Letargat07}
\bibinfo{author}{\bibfnamefont{R.}~\bibnamefont{{Le Targat}}}, Ph.D. thesis,
  \bibinfo{school}{LNE-SYRTE} (\bibinfo{year}{2007}),
  \urlprefix\url{http://tel.archives-ouvertes.fr/tel-00170038}.

\bibitem[{\citenamefont{Slater}(1952)}]{Slater52}
\bibinfo{author}{\bibfnamefont{J.~C.} \bibnamefont{Slater}},
  \bibinfo{journal}{Phys. Rev.} \textbf{\bibinfo{volume}{87}},
  \bibinfo{pages}{807} (\bibinfo{year}{1952}).



\end{thebibliography}
\end{document}